# Coexistence of Scattering Enhancement and Suppression by Plasmonic Cavity Modes in Loaded Dimer Gap-Antennas

*Qiang Zhang,[†] Jun-Jun Xiao,[†,\*] Meili Li,[†] Dezhuan Han,[‡] and Lei Gao[§]*

[†]College of Electronic and Information Engineering, Shenzhen Graduate School, Harbin Institute of Technology, Shenzhen 518055, China

[‡]Department of Applied Physics, Chongqing University, Chongqing 400044, China

[§]Jiangsu Key Laboratory of Thin Films, School of Physical Science and Technology, Soochow University, Suzhou 215006, China

[\*]**Addresses correspondence to eiexiao@hitsz.edu.cn**

**ABSTRACT:** Plasmonic nanoantenna is of promising applications in optical sensing, single-molecular detection, and enhancement of optical nonlinear effect, surface optical spectroscopy, photochemistry, photoemission, photovoltaics, etc. Here we show that in a carefully-designed dimer gap-antenna made by two metallic nanorods, the longitudinal plasmon antenna mode (AM) of bonding dipoles can compete with the transverse plasmonic cavity modes (CMs), yielding dramatically enhanced or suppressed scattering efficiency, depending on the CMs symmetry characteristics (e.g., the radial order $n$ and the azimuthal quantum number $m$). More specifically, it is demonstrated that an appropriately loaded gap layer enables substantial excitation of toroidal moment and its strong interaction with the AM dipole moment, resulting in Fano- or electromagnetically induced transparency (EIT)-like profile in the scattering spectrum. However, for CMs with nonzero azimuthal number, the spectrum features a cumulative signature of the respective AM and CM resonances. We supply both detailed near-field and far-field analysis for these phenomena, showing that the modal overlap and phase relationship between the fundamental moments of different order play a crucial role. A classical coupled oscillator model is proposed which clearly reproduces the coexistence of scattering enhancement and suppression. Finally, we show that the resonance bands of the AM and CMs can be tuned by adjusting the geometry parameters and the permittivity of the load. Our results may be useful in plasmonic cloaking, spin-polarized directional light emission, ultra-sensitive optical sensing, and plasmon-mediated photoluminescence.

**INTRODUCTION**

Plasmonic nanoparticles and their assemblies are well-known optical nanoantennas, and have been intensively studied in nanophotonics due to the fascinating optical properties originated from localized surface plasmon resonance (LSPR).[1-3] Among the various kinds of metallic nanostructures, plasmonic dimer antennas (PDAs) which are often constructed by a pair of strongly interacting metallic nanoparticles, e.g., a dimer consisting of two arms and a gap between them, have been frequently studied.[4-9] Despite the numerous similarities with traditional RF antenna in terms of collecting and emitting electromagnetic wave,[10,11] the LSPRs in PDA offer tremendous extraordinary and anomalous optical responses. More specifically, PDA can confine light in extremely small volume in the gap region[12-14] while it can scatter or absorb light with optical cross sections much larger or smaller than its geometry cross section.[15,16] Moreover, the resonance features of a PDA is intimately related to the size, shape, the dielectric functions of the compositions, and the ambient enviroment.[17-19] The virtues of LSPR grant PDA promising applications in ultrasensitive sensing (e.g., by surface enhanced Raman scattering which has great potential to detect single molecule),[20-22] enhanced light-matter interaction,[23-25] optical nano-circuit,[26-28] enhancement of optical nonlinear effects,[29,30] and plasmon-assisted particle trapping and micromanipulations[31,32] etc. In addition, since PDAs represent one of the simplest coupling systems, they are quite suitable for studying plasmon hybridizations and coherent plasmonic phenomena such as Fano resonance and electromagnetically induced transparency (EIT).[33-37]

To this end, a great deal of attention has been focused on the antenna mode (AM) of a PDA which is basically an electric dipolar plasmon resonance sustained by the solid metallic parts (e.g., the antenna arm). The nanogap of a PDA provides the feeding port to excite the AM, usually via a transmission line or a local emitter. In the same time it represents a load to tune the equivalent circuit property (e.g., the input impedance).[15,18,38] However, in such plasmonic nanostructures, not only the solid metallic parts but also their inverse counterparts (e.g., the dielectric gap layer embedded between the metals) play significant roles. The LSPRs associated with the solid metallic parts usually give rise to prominent electric resonances while the inverse parts favor magnetic resonances, according to the Babinet principle.[39-41] It is more fundamental to realize that PDAs are actually composite structures with the arms being the solid part and the dielectric gap being the inverse part, simultaneously. Indeed, when the PDA has extremely short arms, it becomes a patch metallic-dielectric-metallic (MDM) antenna that has been studied extensively.[42-47] Circular patch MDM antennas sustain a series of magnetic cavity modes (CMs) which are excitable by the magnetic component of impinging light, and these CM modes are perceived as transverse plasmonic standing waves inside the dielectric gap layer.[42]

Different to the AM, the CMs are strongly confined in the dielectric layer with small modal volume and high quality factor.[43,44] In view of the multipole expansions, the lower-order CMs give rise to the fundamental magnetic multipoles.[46,47] More interesting, one of the CMs has been shown to generate remarkable toroidal dipole response. The toroidal dipole response could yield interesting consequences such as the formation of anapole,[48] toroidal induced transparency,[49-51] and enhanced optical scattering force.[52] In this context, one would expect that a PDA may support both AM and CMs in the same band and it shall be of great interest to explore the couplings between them. We note that they were respectively analyzed in detail in a very recent work.[53] However, studies on the AM-CM coupling effects are still missing. As a matter of fact, there are several reasons behind the overlook of these couplings: (i) commonly used PDAs are simply of air gap and of large aspect ratio, resulting in far separation of the AM and the CM resonance frequencies: The former generally occurs at relatively low frequency and the latter lies at relatively high frequency in most cases; (ii) in extremely thin patch MDM antennas, the CMs are tuned to relatively low frequency while that of the AM is in very high frequency, even beyond the bulk plasmon frequency and non-survivable; (iii) the quality factor of the AM is usually much smaller than those of the CMs, making the latter hard to excite. Therefore the signatures of CMs in the optical spectrum are substantially submerged in a relatively broadband dipolar resonance background. This is further deteriorated when the gap layer is much thinner than the length of the antenna arm.

Here in this study, we overcome these issues by carefully adjusting the geometry and the gap layer material in a PDA consisting of two identical silver nanorods. By deliberately making the AM and CMs spectrally overlapped, we are able to simultaneously or selectively obtain destructive and cumulative responses from them, depending on the angular symmetry of the CMs. Both far-field and near-field characteristics of the designed PDAs are examined by full-wave simulations based on the finite integral technique (FIT).[54] The results show that the magnetic dipolar CM fully decouples from the electric bonding dipolar AM and they collectively generate accumulative scattering enhancement. On the contrary, the toroidal dipolar CM can violently compete with the AM in the near field, yielding Fano resonance or EIT-like features in the spectrum. It should be noted that both the toroidal CM and the dipolar AM involves zero angular momentum in the electric field distribution. The distinct consequences by the magnetic dipolar CM and the toroidal dipolar CM can be understood in terms of modal overlap integral and their orthogonality with respect to the AM. A classical coupled oscillator model is then established to understand the underlying interactions and reproduces the results qualitatively. It shall be stressed that the Fano or EIT-like phenomena comes from the near-field competition in the gap region rather than from far-field scattering interference, in contrast to the cases as shown by Liu et al.[49,50] To the best of our knowledge, this work represents the first proposal of strong EIT-like scattering phenomena produced by the coupling between AM and CM in a plasmonic homodimer antenna.

**RESULTS AND DISCUSSIONS**

Figure 1a schematically shows the PDA structure which is a typical homodimer with two identical silver nanorods loaded with a dielectric layer of thickness $d$ inside the gap. The two circular nanorods are both of radius $R$ and length $L$. The dielectric constant of the gap material is set to be $\varepsilon_{load}$ and that of the silver is taken from Johnson and Christy.[55] The whole structure is assumed to be freestanding in air and illuminated by a normally incident plane wave ($\mathbf{k}$ along $+\hat{z}$ direction) that is linearly polarized along the $x$ axis. Firstly, we set $R = 50$ nm, $L = 90$ nm, $d = 20$ nm and check how the scattering spectrum evolves with various $\varepsilon_{load}$. The calculated optical scattering efficiency $\sigma_{scs}$ for $\varepsilon_{load} = 1$, 3, 6, 12 and 18 are shown in Figures 1b-1e. It is seen that for $\varepsilon_{load} = 1$ the spectrum has a broad peak centered at $f \approx 530$ THz (see Figure 1b). This peak corresponds to the electric dipolar AM resonance which is also known as the bonding hybridization resonance from the individual dipolar modes on the two nanorods.[35] Figure 2c shows that as the gap dielectric constant increases to $\varepsilon_{load} = 3$, a narrow Lorentz-shape peak (marked by the red upward arrow) is superimposed on the AM peak which red shifts to $f = 462$ THz. With loaded dielectric material of $\varepsilon_{load} = 6$, it is seen in Figure 1d that the narrow peak ($f = 358$ THz) resides at the left shoulder of the broad AM resonance. More interestingly, another resonance feature shows up near $f = 540$ THz, apparently with a quite different flavor. This resonance reduces rather than adds up to the AM resonance profile. It also leads to an asymmetric line shape with a Fano-like dip, as marked by the downward blue arrow in Figure 1d. Following this trend, when $\varepsilon_{load}$ is continuously increased to 12, which is the situation shown in

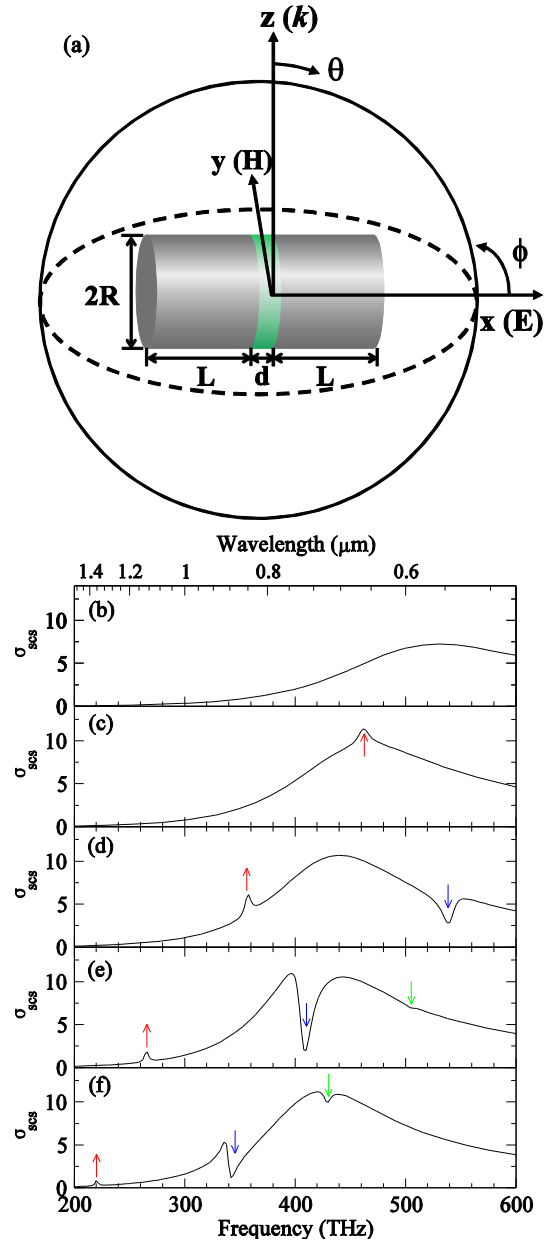

**Figure 1.** (a) Schematic figure of the PDA made of two identical silver nanorods and a dielectric load in the gap region. The nanorods have radius $R$ and length $L$, while the thickness of the gap layer is $d$. The incident plane wave propagates along $+\hat{z}$ direciton with polarization along the $x$-axis. The scattering efficiency spectrum $\sigma_{SCS}$ of a PDA with $L = 90$ nm, $R = 50$ nm, and $d = 20$ nm for (b) $\varepsilon_{load} = 1$, (c) $\varepsilon_{load} = 3$, (d) $\varepsilon_{load} = 6$, (e) $\varepsilon_{load} = 12$, and (f) $\varepsilon_{load} = 18$. The scattering peaks in (c)-(f) are marked by upward arrows, and the first and second Fano dips are marked by downward blue arrows and downward green arrows, respectively.

Figure 1e, both the upward-arrow-marked peak and downward-arrow-marked dip red shift. However, their relative magnitudes with respect to the AM background (envelope) exhibit different changing characteristics. The strength of the Lorentzian peak nearly does not change while that of the Fano dip visually increases as it approaches the resonance center frequency of the AM. It is seen that the original AM peak becomes a deep EIT-like dip and splits into two separate peaks at $f = 398$ THz and 444 THz, respectively. This indicates a strong interaction of the AM with another resonance mode (we will later show that it is a CM with toroidal response). Notice that in Figure 1e, there is another tiny Fano dip at $f \approx 505$ THz (marked by the downward green arrow) which must come from a higher-order mode. Finally, Figure 1f shows that for $\varepsilon_{load} = 18$, this high-energy dip looks apparent as it now spectrally approaches the AM resonance center frequency.

It is reasonably to infer from the above observations that the sub-features in the spectra, e.g., the peaks and the Fano-like dips, may come from another family of resonances that are distinct from the prominent AM. To clarify that, we focus on the PDA structure studied in Figure 1e. Figure 2a shows the absorption efficiency $\sigma_{ACS}$ together with the scattering efficiency $\sigma_{SCS}$ of the whole structure. The FIT results were further corroborated by a solver (COMSOL Multiphysics) of finite element method (FEM).[56] It is seen that the FEM results (symbol) agree very well with the FIT ones (red line). The $\sigma_{ACS}$ spectrum (dashed line) exhibits three peaks at $f = 258$ THz, 408 THz and 505 THz, with diminishing amplitude in sequence. With regard to the low-energy $\sigma_{ACS}$ one at $f = 258$ THz, it coincides with the $\sigma_{SCS}$ peak. However, for the $\sigma_{ACS}$ peak at $f = 408$ THz and 505 THz, they are very close to the corresponding $\sigma_{SCS}$ dip.

To identify these resonances, we show in Figures 2b-2f the electric field amplitude $|\mathbf{E}|$ over the $xoz$ plane for $f = 265$ THz, 398 THz, 409 THz, 444THz and 505 THz, respectively. It is seen that at $f = 265$ THz, the electric field is basically confined in the gap region with a zero-field node in the center (see Figure 2b). This strongly suggests that it originates from a plasmonic cavity mode. Figure 2h shows the out-of-plane electric field $E_z$ (see the color contour) in the dielectric layer across the origin. We label this resonance as 'CM$_{11}$' where the subscript represents the radial and azimuthal node numbers in the $E_z$ pattern. Furthermore, the $yoz$ in-plane magnetic field vectors are shown by arrows which exhibit a large magnetic moment along the $y$ direction. This magnetic moment is formed by the anti-directional going conduction currents on the opposite gap surfaces, given by the so called magnetic resonance.[42]

Figure 2c shows that for the dominated scattering peak at $f = 398$ THz at the lower-frequency side of the Fano dip, there are hot spots both inside the gap region and near the antenna terminals. Strong electric field around the antenna terminals is exactly a signature of the AM. In fact, here it is identified as an electric dipolar resonance that strongly radiates, despite of the slight asymmetric characteristic along the $z$ axis due to retardation. At the same time, Figure 2c shows that the electric field is also much stronger inside the gap, with the zero-field node appearing near the lateral boundary. Figure 2i indicates that at this frequency the PDA also sustains the CM$_{10}$ resonance in the gap region. These imply hybridization between the dipolar AM and the cavity mode CM$_{10}$. For convenience, here we label the peak by 'AM$_1$' to distinguish it from the other hybridized mode 'AM$_2$' that contributes to the second overwhelming scattering peak at $f = 444$ THz. Figure 2d shows the electric field pattern for the scattering dip at $f = 409$ THz that falls in between AM$_1$ and AM$_2$. Obviously, the field around the antenna terminals supposedly produced by the AM nearly vanishes (see Figure 2d). We note that this vanishing AM field highlights one of the important features of Fano resonance: the "bright" mode excitation is substantially suppressed at the Fano dip frequency.[36] In a similar manner, we label this resonance as 'CM$_{10}$' according to the $E_z$ field pattern shown in Figure 2j wherein the in-plane magnetic field is in a circular form confined in the gap region. Such magnetic field confinement gives rise to a toroidal dipole response, as observed in circular patch MDM antennas.[46,47] For frequency $f = 444$ THz beyond the Fano dip, the scattering peak from 'AM$_2$' emerges. Figure 2e shows that the hot spots around the antenna terminals recover to some extent (the frequency is away from the AM resonance center frequency) but those inside the gap diminish. The near field maps for the high-energy minor dip at $f \approx 505$ THz are shown in Figures 2f and 2l. Both of them help to confirm that the gap mode at this frequency is 'CM$_{20}$'.

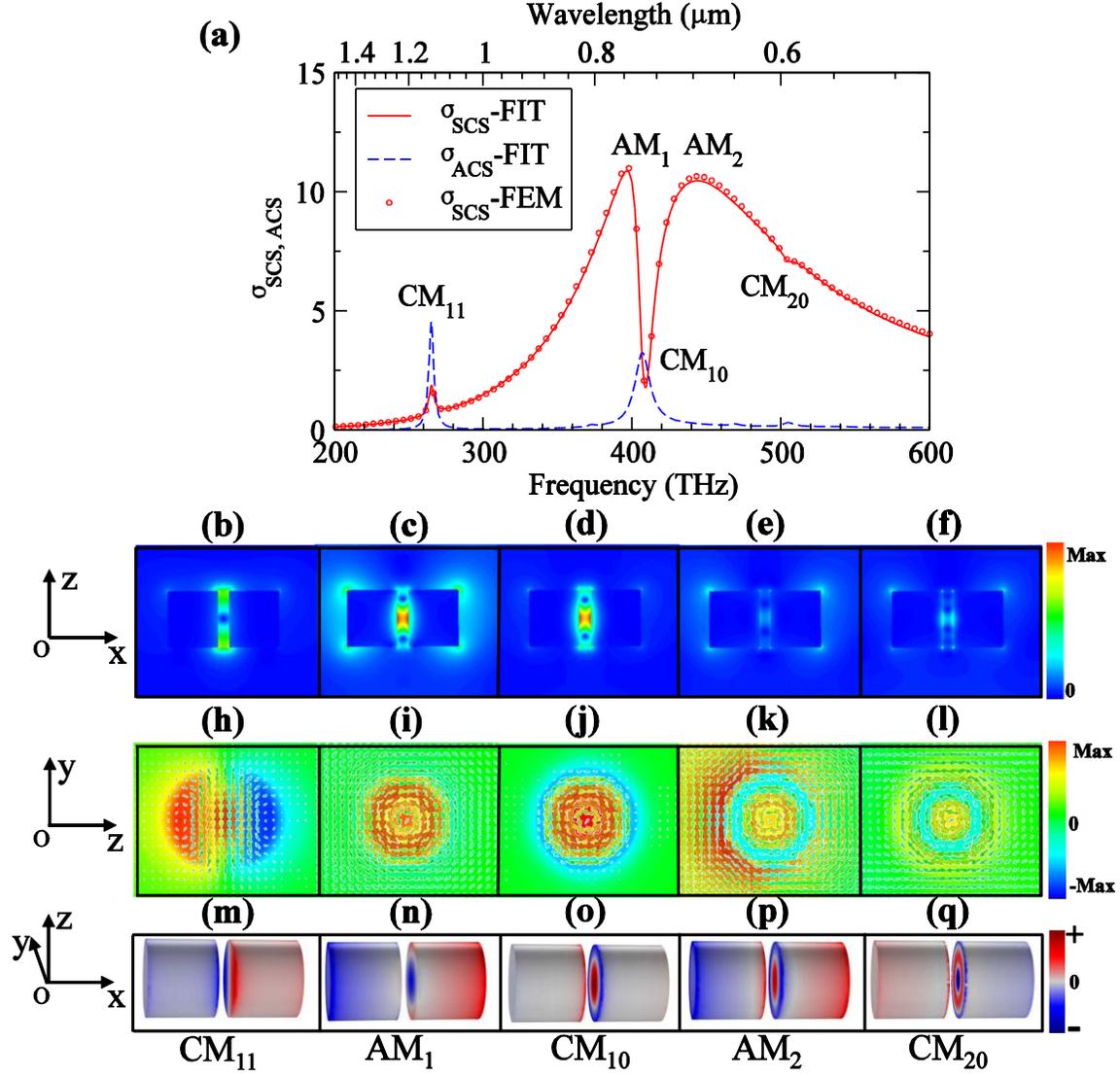

**Figure 2.** (a) The $\sigma_{SCS}$ (red line) and $\sigma_{ACS}$ (blue dashed line) spectra of the PDA with $L = 90$ nm, $R = 50$ nm, $d = 20$ nm and $\varepsilon_{load} = 12$. The red dots are results calculated by the FEM. Three significant CMs features at $f = 258$ Hz, 409 THz and 502 THz are labeled as 'CM$_{11}$', 'CM$_{10}$' and 'CM$_{20}$' respectively. The two peaks at $f = 398$ THz and 444 THz are labeled as 'AM$_1$' and 'AM$_2$'. (b)-(f) The electric field amplitude $|\mathbf{E}|$ at $xoz$ plane for CM$_{11}$, AM$_1$, CM$_{10}$, AM$_2$ and CM$_{20}$ in order. (h)-(l) The out-of-plane electric fields $E_z$ (colors) and the in-plane magnetic field (arrows) at the $yoz$ plane across the dielectric layer. (m-q) The corresponding surface charge distribution.

The above observations show that the AM-CM interaction crucially depends on the CMs' properties. To see that in more detail, we additionally examined the resonance-induced surface charges. Figures 2m-2q plot the surface charge distribution for the labeled peak and dip frequencies in Figure 2a. Overall, the induced charges at the two arms of the PDA are basically of opposite sign, simply because the frequency band from $f = 200$ THz—600 THz is covered by the bonding dipolar AM resonance. However, the fine structures of the charge distributions on the opposing interfaces across the gap are completely determined by the respective CM resonance. For example, Figure 2m shows that for CM$_{11}$, the positive and negative charges on the two inner interfaces across the gap separate along the $y$ axis, with a neutral node around the origin. Meanwhile, Figures 2n-2o show that for AM$_1$, CM$_{10}$, and AM$_2$, the induced opposite charges around the gap distribute alternatively in the radial direction without azimuthal zero node. We note that this "positive-negative" ring-like charge distribution is very similar to the reported fine structure in nanosphere heterodimers at the Fano dip frequency.[34] Such distribution to some extent is ascribed to the cavity in between. Indeed, the suppression of the AM at the Fano dip is more clearly seen in Figure 2o where the metallic arms are mostly neutralized in the terminals and the surface charges accumulate near the gap interfaces. Figure 2q shows that for the $CM_{20}$ resonance, the charges near the gap have connected ring-like pattern.

As a matter of fact, there are several other CMs in this frequency band with undistinguishable peak in the $\sigma_{SCS}$ spectrum due to the weak excitation (see Supporting Information, Figure S1). To make our discussions more concise and simple, hereafter, we ignore them and the $CM_{20}$ mode, and specifically focus on the $CM_{11}$ and $CM_{10}$ modes. Figures 2h and 2j demonstrate that the $CM_{11}$ and the $CM_{10}$ have quite different symmetry pattern. It is thus expected that their modal overlapping with the AM must be distinct, yielding weak and/or strong mutual interactions, respectively. Note that a pure dipolar AM has relatively homogenous electric field in the gap region, with approximately zero azimuthal quantum number ($m=0$). As a consequence, the field overlap integral between $CM_{nm}$ and AM is determined by $\int_0^{2\pi} e^{i(m-m')\phi}d\phi = 2\pi\delta(m-m')$. It is immediately clear that CMs with $m=0$ have nonzero modal overlapping with the AM, otherwise they are orthogonal to the AM. This argument holds as long as both the AM and CMs have negligible in-plane electric field.

The AM and CMs are both excitable in the configuration of Figure 1a since the incoming plan wave simultaneously provides electric component along the axis of the arms and magnetic component across the gap. In other configurations, only one family of plasmon AM and CM can be observed in the scattering spectra. The same features can be seen in the radiation decay rate of an electric dipole source (see Supporting Information, Figure S2).

It is now unambiguous that the $CM_{11}$ and the $CM_{10}$ have quite different interaction with the AM. The $CM_{11}$ mode shows no considerable effect on the AM, but simply presents a cumulative peak in the scattering spectrum (Figure 1). On the other hand, the $CM_{10}$ mode interacts strongly with the AM, leading to a Fano (or EIT-like) dip (see Figure 2a). More importantly, the interaction strongly suppresses the AM in the near fields due to destructive competition on the surface charge distribution, as demonstrated in Figures 2d and 2o. To gain deeper understanding of the underlying physics, we have performed multipole decompositions of the total scattering fields for the PDA structure in Figure 2. Figure 3a shows the scattering efficiency from the irreducible Cartesian multipoles that include the electric dipole (ED), the magnetic dipole (MD), the toroidal dipole (TD), the ED-TD cross term, and the electric quadrupole (EQ) and magnetic quadrupole (MQ).[48-52] It is seen that the scattering efficiency is dominantly contributed by the ED (see the red line). The ED is of no doubt coming from the bonding electric dipolar AM which is relatively bright, namely efficiently excitable and radiating. More importantly, at the $CM_{10}$ resonance $f = 409$ THz, the scattering efficiency from ED shows a dip, indicating that the AM is somehow suppressed. The scattering efficiency of the MD (see the blue line in Figure 3a) is much smaller than that of the ED, with an exception near $f = 265$ THz. This is strong evidence that the $CM_{11}$ mode has a MD response, consistent with the preceding discussions. Different to the MD, the TD scattering efficiency (green line in Figure 3a) in the whole frequency band is at least two orders of magnitude smaller than that from the ED, even near the resonance frequency of $CM_{10}$ ($f = 409$ THz). We note that in most cases the scattering ability of a TD is indeed much weaker than an ED.[51,52] Figure 3a further shows the scattering efficiency from several higher-order mutipoles. For example, the contribution from EQ (red dashed line), MQ (blue dashed line), and the ED-TD cross term notated as '**P**∗**T**' (see magenta solid line). Although the scattering efficiency from EQ and MQ is relatively negligible, they provide extra information. For instance, there is in fact a CM resonance showing MQ response at $f \approx 374$ THz (see the blue dashed line in Figure 3a and Figure S1 in the Supporting Information).

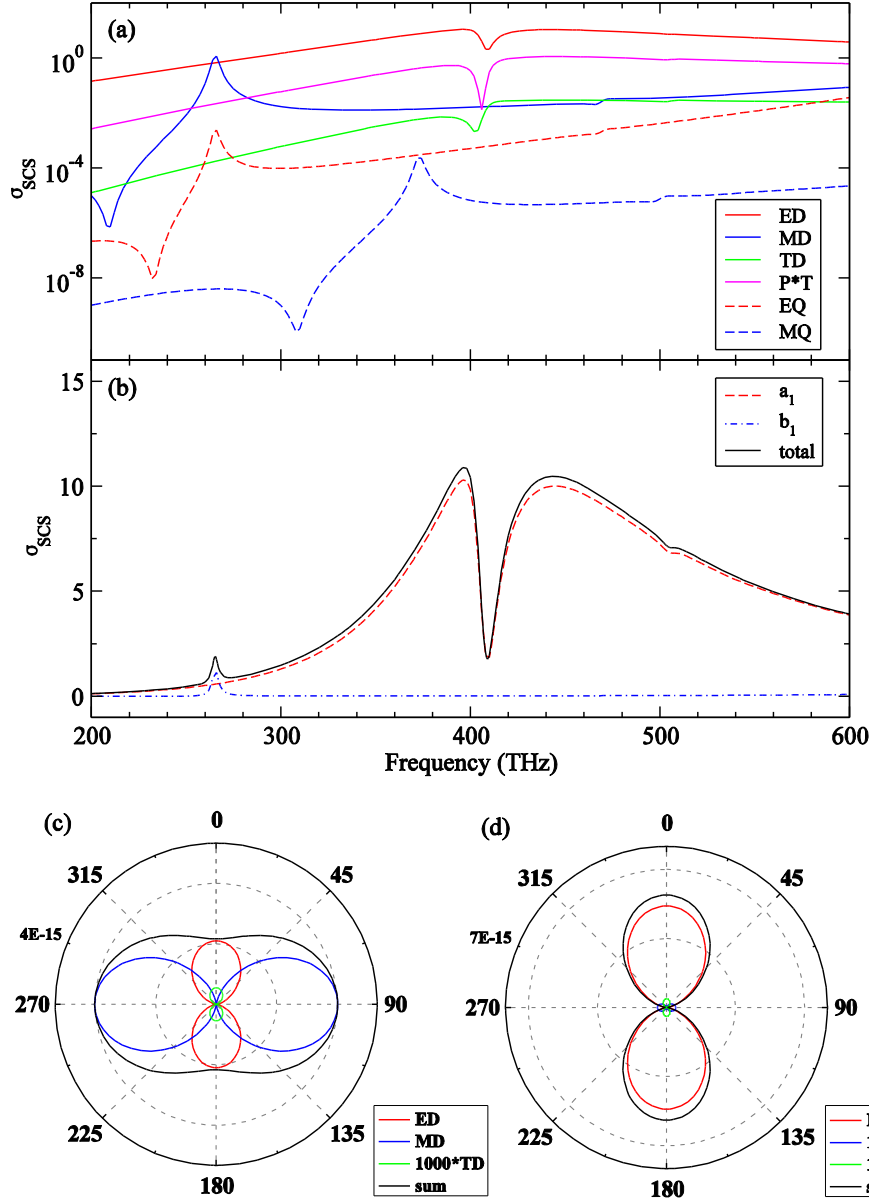

**Figure 3.** (a) The scattering efficiency $\sigma_{SCS}$ of the irreducible multipoles in the Cartesian coordinates. (b) The scattering efficiency $\sigma_{SCS}$ of the spherical dipole $a_1$ and $b_1$ and the total $\sigma_{SCS}$ which is the same as the red line in Figure 2(a). (c) The radiation pattern of ED, MD, TD and their vectorial summation in the $\theta = 0$ plane for the $CM_{11}$ resonance at $f = 258$ THz. Note that the far field intensity from the TD is artificially amplified 1000 times to increase the visibility. (d) Same as (c) but for the $CM_{10}$-induced Fano dip at $f = 409$ THz. The TD term is multiplied by 10.

Figure 3b shows the decomposed scattering spectra from multipole moments in spherical coordinates (see METHODS) and the total $\sigma_{SCS}$ (solid black line). The spherical scattering coefficients $a_1$ and $b_1$ are the main contributions: their sum basically recovers the total $\sigma_{SCS}$. We would like to emphasize that $a_1$ contains both the ED and TD parts in Figure 3a.[53] We further note that the scattering dip in the $\sigma_{SCS}$ spectrum is essentially different to the recently reported toroidal induced transparency which is basically ascribed to the scattering cancellation between ED and TD.[48-51] The scattering cancellation demands comparable strength of the individual scattered powers from both the ED and TD, which is not the case in our system. Figure 3a confirms that the TD scatters much more weakly (two orders of magnitude lower) as compared to the ED. In view of that, we would ascribe the observed scattering dip here to Fano suppression induced by the strong near field competition (mainly inside the gap region) between the AM and the $CM_{10}$ mode.

In addition to the scattering properties shown in Figures 3a and 3b, the radiation pattern can present more angular and directional information. Figures 3c and 3d show the far field intensity on the plane of $\theta = 90°$, at $f = 258$ THz (the magnetic resonance) and $f = 409$ THz (the Fano dip), respectively. The polar plots compare contributions from the ED (red line), MD (blue line), TD (green line) and their vectorial summation (black line). Notice that in Figure 3c the far field strength of TD is artificially amplified by 1000 times. Both radiation patterns have typical dipolar form but with different strength and orientation feature. Primarily, the scattering patterns of the ED and TD have the same angular momentum since they are both parallel to the electric component (along the $z$ axis) of the incident wave and belong to the spherical dipole $a_1$.[48-52] Regarding the scattering pattern of the MD, it is rotated by 90° with respect to that of ED and MD. The MD is parallel with the magnetic component (along the $y$ axis) and belongs to the spherical dipole $b_1$.[57] The orthogonal ED and MD are crucial for unidirectional scattering[58-60] and spin-dependent photon emissions.[61] Here the designed PDA loaded with $\varepsilon_{load} = 3$ also exhibits the backward scattering suppression at the CM$_{11}$ resonance (Supporting Information Figure S3).

Figure 3 shows that both the ED and MD contribute to the scattered field with different strength and channel. Therefore, it is reasonable to consider the AM as a bright mode and the CM$_{11}$ mode as a sub-bright mode. Concomitantly, the CM$_{10}$ mode hardly radiates and can be considered as a dark mode that only strongly interacts with the AM in the near field. In view of that, we employ a classical coupled oscillator model to qualitatively understand the observed scattering spectrum. Figure 4a sketches the scheme where $G_p$ ($G_m$) represents the coupling of AM (CM$_{11}$) to the external drive, $\kappa$ being the coupling strength between the AM and CM$_{10}$. The energy spectrum of each oscillator is schematically shown in the right side of Figure 4a. The dynamic equations of this system follows[62]

$$\begin{pmatrix} \chi_p \\ \chi_m \\ \chi_t \end{pmatrix} = \begin{pmatrix} \omega_p^2 - \omega^2 + i\gamma_p\omega & 0 & \kappa^2 \\ 0 & \omega_m^2 - \omega^2 + i\gamma_m\omega & 0 \\ \kappa^2 & 0 & \omega_t^2 - \omega^2 + i\gamma_t\omega \end{pmatrix}^{-1} \cdot \begin{pmatrix} G_p \\ G_m \\ 0 \end{pmatrix} \quad (1)$$

where $\chi_p$, $\chi_m$ and $\chi_t$ denote the responses of the bright, the sub-bright and the dark modes, respectively. The three modes have eigenfrequency $\omega_p$, $\omega_m$ and $\omega_t$ and intrinsic dissipation $\gamma_p$, $\gamma_m$ and $\gamma_t$, respectively. In Equation (1) we have set zero coupling coefficients between $\chi_m$ and $\chi_p$ ($\chi_t$) since they are spectrally far away and considered to be nearly orthogonal. The extinction spectrum of the whole coupled system is measured by $-\omega \operatorname{Im}(\chi_p + \chi_m)$ which accounts the total dissipated powers. Figure 4b shows that the model results agree qualitatively with the numerically calculated extinction spectrum $\sigma_{ECS}$. Despite of the small discrepancy between the model (line) and the FIT numerical results (circles), it is seen that the model essentially captures the features of the coexistence of scattering enhancement and suppression.

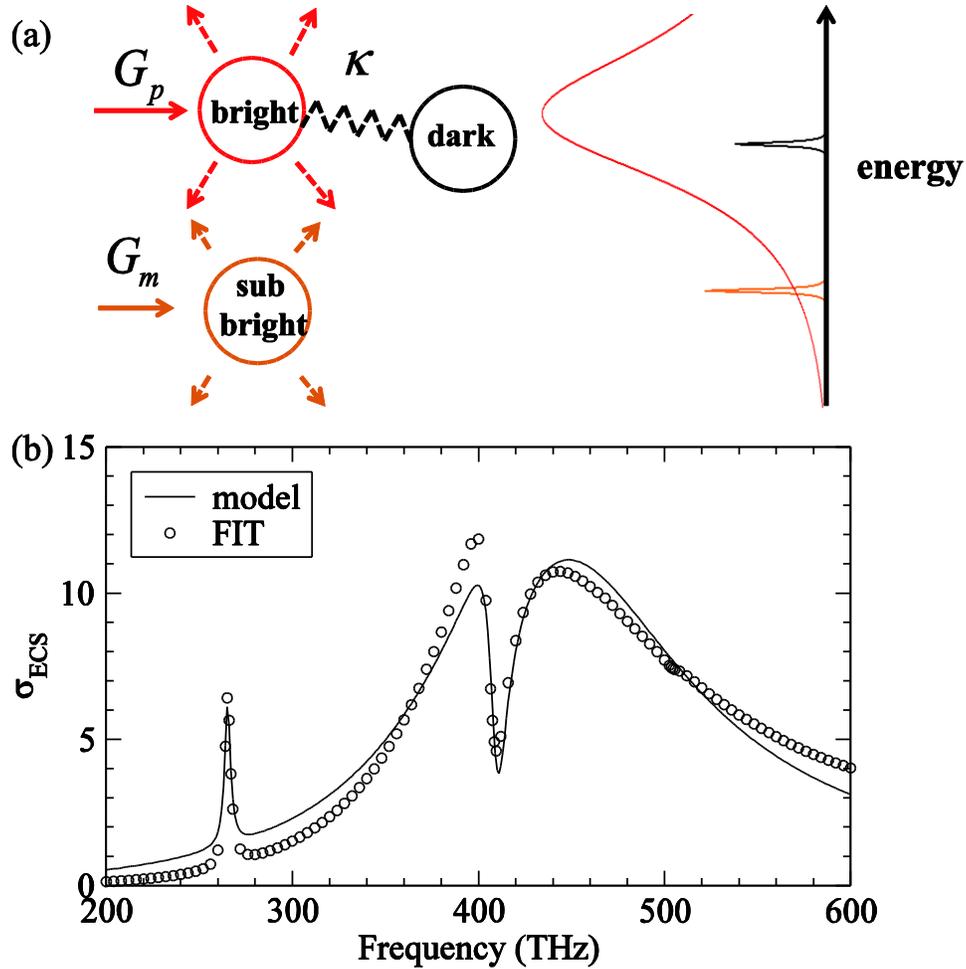

**Figure 4.** (a) The coupled oscillator model for the AM-CM interactions. The AM is considered as the bright mode (red circle) with excitation strength $G_p$, the $CM_{11}$ being a sub-bright mode (orange circle) with excitation strength $G_m$, and $CM_{10}$ the complete dark mode (black circle). The coupling coefficient between the AM and $CM_{10}$ is $\kappa$. Their individual resonance spectra are schematically shown at the right side. (b) The $\sigma_{ECS}$ spectrum of the PDA numerically calculated by the FIT (circles) and fitted result from the coupled oscillator model (line).

It should be stressed that both modal field overlap (nonzero inner product of eigenmodes) and spectral overlap are necessary prerequisites to guarantee strong coupling between the AM and CM modes. In this regard, it is interesting to see how the scattering spectra are affected by these factors in realistic design. Figures 5a-5d show the $\sigma_{SCS}$ contour maps in the frequency band from $f = 200$ THz to $600$ THz with varying $\varepsilon_{load}$, $d$, $L$ and $R$, respectively. The default parameters are kept the same as those in Figure 2. Firstly we see that all the $\sigma_{SCS}$ maps exhibit a dominating broad band coming from the AM resonance. There are also three visible narrow bands from the $CM_{11}$, $CM_{10}$ and $CM_{20}$ resonances. It is clear that the $CM_{11}$ band is a scattering enhanced one and the $CM_{10}$ and $CM_{20}$ bands correspond to two Fano bands with suppressed scattering. More interestingly, the strong Fano dip induced by the $CM_{10}$ cuts through the dominating AM band in the entire frequency range. It thus introduces an anti-crossing feature, reflecting the strong interaction between the AM and the $CM_{10}$ modes. The resonance frequencies of the CMs are theoretically predictable based on a Fabry-Perot model,[53] or approximately by applying the Neumann boundary condition for cylindrical gap SPPs at the lateral interface between the gap layer and the exterior medium for a circular MDM cavity.[42,45] The principle of this method is based on the fact that the resonance frequency of $CM_{nm}$ versus $\chi'_{mn}/R_{eff}$ shall approximately fall on the dispersion of the gap SPPs, where $\chi'_{nm}$ is the $n$-th zero point of the derivative of $m$-th Bessel function.[42,45,63] Therefore the resonance frequency of the CMs are mainly determined by two factors: (1) the effective radius $R_{eff}$ and (2) the wave vector of the gap SPP $k_{gsp}$, both of which are closely related to the geometry and loaded material $\varepsilon_{load}$. The design parameter dependent resonance frequencies of the $CM_{11}$, $CM_{10}$ and $CM_{20}$ cavity modes are theoretically obtained and shown in Figure 5 (black dashed lines). The theoretical prediction clearly follows the local maximum and minimum of the color contour.

In more details, Figure 5a shows that as the loaded material dielectric function increases from $\varepsilon_{load} = 1$ to $21$, both the AM band and the CM bands redshift. However, the AM band is less sensitive than the CM bands with respect to the $\varepsilon_{load}$ variation. In particular, for roughly $\varepsilon_{load} \geq 9$, the AM band becomes flat at fixed frequency range from $f = 400$ THz to $500$ THz. Thus it is possible to adjust the CM bands independently by altering the loaded material. Figure 5a further shows that for air load, e.g., with $\varepsilon_{load} = 1$, the resonance frequency of CMs is beyond $f = 600$ THz which is far from the AM band. This is one of the reasons for the absence of the AM-CM coupling effect in the literature. The red-shift of the CM bands is caused by the increased propagating constant $k_{gsp}$ in the high dielectric constant layer. With regard to the AM, Alú et al have demonstrated that the load can be applied to tune the scattering properties of the total antenna.[15] The optical capacitance of the dielectric load is approximately $C_{load} = \varepsilon_{load} \pi R^2 / d$.[15,18,64] In this view, the red-shift of the AM band is a straightforward result from the resonance condition $\omega_0 \propto \sqrt{1/C_{load}}$. However, this argument is valid only when we are dealing with a pure electric dipolar mode. As long as the AM is hybridized with the CMs, the situation becomes more complicated.

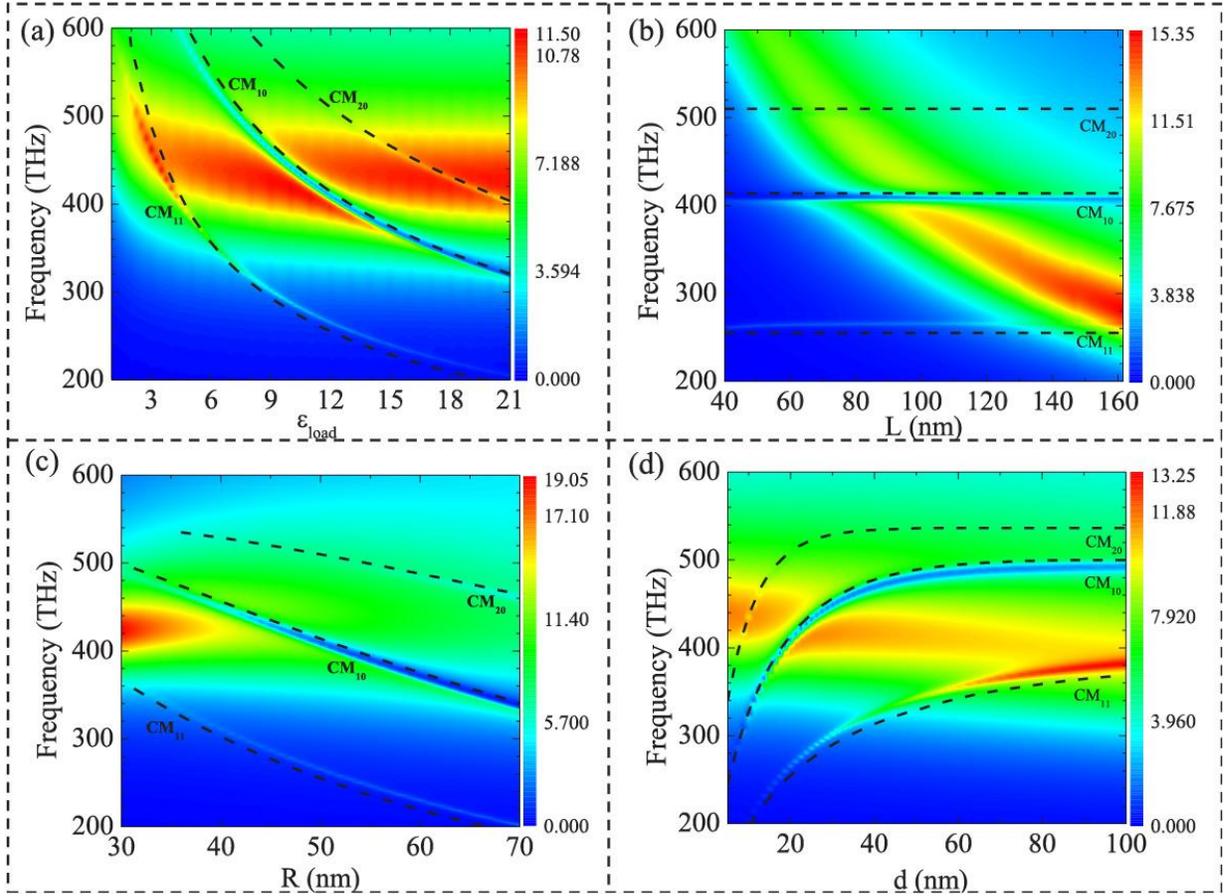

**Figure 5.** The scattering efficiency $\sigma_{SCS}$ contour in the frequency—parameter space. (a) $\varepsilon_{load}$ from 1 to 21. (b) $L$ from 40 nm to 160 nm. (c) $R$ from 30 nm to 70 nm. (d) $d$ from 5 nm to 100 nm. The default parameter values are the same in Figure 2. The dashed black lines show the frequency dependence on the respective design paramaters. They are obtained by theoretical prediction for the $CM_{11}$, $CM_{10}$ and $CM_{20}$ resonances.

It is probably more feasible to tune the scattering of the antenna by adjusting its geometry parameters. Figure 5b plots the $\sigma_{SCS}$ map for arm length $L = 40$ nm to 160 nm. It is seen that the AM band red-shifts rapidly while the CM bands stay nearly intact. This means that modulating the length of the metallic nanorod is an effective way to tune the AM band without affecting the CM bands much. We note that the CM bands also shift when $L$ is very small because in that scenario the SPPs at the external interfaces of the metallic arms couple with the gap SPPs.[42] The CM bands are more sensitive than the AM band with respect to the radius variation, as shown in Figure 5c. Gap cavity with bigger size favors gap SPP resonances at longer wavelength. It is also worth mentioning that the magnitude of $\sigma_{SCS}$ falls quickly as $R$ increases due to the reduced aspect ratio $L/R$. Figure 5d depicts the $\sigma_{SCS}$ map as a function of the gap distance $d$. It is known that $k_{gsp} \sim \omega$ curve descends and finally converges to that of the SPP at a metal-dielectric interface when $d$ is large enough. As a result, the CM bands are blue-shifted and nearly saturate when $d \geq 60$ nm for $CM_{10}$ and $CM_{20}$ bands. Figure 5d further shows that the AM bands slightly red shift as $d$ increases. This is inconsistent to both the nano-circuit theory[38] and the dipole plasmon ruler equation[65] that have blue shifting prediction for increasing $d$. We argue that these long-wavelength limit theories are valid for solid metallic particles solely supporting electric modes, which is not the case in our system. In the proposed PDAs, the AM is no longer a pure electric mode since it strongly couples to the $CM_{10}$ mode and also couples to the $CM_{20}$ mode. As a matter of fact, this cavity-mediation-effect was recently reported in a nancube dimer with extremely short separation.[66] The failure of the dipole approximation at shot separation is ascribed to the inhomogeneous field induced by the cavity mode, similar to those studied here. As the $CM_{10}$ band is far from the AM band, the AM center resonance frequency blue shifts for increasing $d$. (Supporting Information, Figure S4)

**CONCLUSION**

In summary, we have systematically explored some anomalous scattering properties of nanoantennas made by silver nanorod homodimer. The antennas are deliberately designed to support spectrally overlapping longitudinal electric dipolar AM and transverse gap SPP CMs, by carefully selecting the gap medium and the nanorod geometry. It is found that the broad scattering peak of the AM is modulated by a cumulative narrow scattering peak and one or several Fano dips, depending on the CM's characteristics (e.g., the radial order $n$ and the azimuthal quantum number $m$). More specifically, it is demonstrated that an appropriate gap layer enables substantial excitation of toroidal moment and its strong interaction with the AM dipole moment, resulting in Fano- or EIT-like profile in the scattering spectrum. By analyzing the near field characteristics carefully, we confirm that the narrow peak is related to the $CM_{11}$ resonance while the Fano dip corresponds to the $CM_{10}$ resonance. The excitation of the CMs results in fine structures in the local fields and the surface charge distributions across the gap. Multipole decomposition analysis shows that the $CM_{11}$ mode generates remarkable magnetic dipole moment while the $CM_{10}$ has unnegligible toroidal dipole response. The multipole expansions also reveal that scattering properties of the $CM_{11}$ and $CM_{10}$ are quite different in both amplitude and directional pattern. Despite of the comparable scattering intensity of the $CM_{11}$ with respect to that of the dominated PDA, they scatter into different angular momentum channels. On the other hand, the $CM_{10}$ weakly radiates but strongly interacts with the AM in the near field, producing violent Fano resonance. A coupled system of bright mode, sub-bright mode and a dark mode reproduces the numerically results qualitatively, offering a classical and intuitive picture for the underlying physics.

We have also studied how the $\sigma_{SCS}$ maps evolve by tuning the load and the geometry parameters of the antenna. It is possible to tune the AM band and the CM bands independently, achieving various type of spectrum. We believe the porous anodized alumina (AAO) template is probably an effective platform to fabricate the proposed structures. The silver—dielectric—silver layers can be sequently eletrodeposited into the pores of a well-prepared AAO template.[6,8] In experiments, the optical scattering spectra reported here should be observable by a dark-field spectroscopy. As we have shown, the whole spectrum can be tuned by a bunch of geometry parameters and also by the gap layer properties, dynamic tuning, e.g.,by electro-optic, magneto-optic, and thermo-optic approaches, is possible. We stress that the generation of Fano resonance in nanostructures often demands symmetry broken such as using heterodimers,[34-36] oblique excitations, or gyromagnetic substrates etc.,[67] to invoke the dark mode. Here we present a new way to induce Fano resonance by coupling the AM and CMs in a symmetry optical antenna. In particular, we show that a toroidal moment could crucially compete with an electric dipole moment in the near field. Of course, more coherent phenomena are expected by involving symmetry broken, such as using a heterodimer, accounting possible substrate effects, using anisotropic gap layer, and so on. Our findings provide more degree of freedom to manipulate the resonances in PDA, by utilizing both the AM and the CMs. The results presented here may find applications including plasmonic superscattering and cloaking, optical sensing based on Fano resonance, spin-dependent and directional light emission, and efficient light harvest employing optical antennas.

**METHODS**

**Scattering cross section calculation.**

In the FIT simulations (CST Microwave Studio), the PDA is placed in vacuum with open add space boundary conditions. We have used the adaptive refinement tetrahedral meshes in the frequency domain solver. By pre-defining the far field monitors, $\sigma_{SCS}$ and $\sigma_{ACS}$ are calculated automatically. In order to guarantee the accuracy, the main results are also checked by employing the FEM simulation (COMSOL Multiphysics). In the FEM simulations, perfectly matched layer (PML) boundary conditions are used. The $\sigma_{SCS}$ spectrum are obtained by integrating the normal scattered Poynting over a spherical surface enclosing the PDA. In the same time, the $\sigma_{ACS}$ spectrum are obtained by integrating the resistive losses over the PDA volume. The extinction is defined by their sum $\sigma_{ECS} = \sigma_{SCS} + \sigma_{ACS}$.

For all the numerical simulations, the permittivity of silver $\varepsilon(\omega)$ is taken from the experimental results of Johson and Chrisity fitted by the following Drude-Lorentz model:

$$\varepsilon(\omega) = 1 + \varepsilon_\infty - \frac{\omega_p^2}{\omega(\omega + i\Gamma_0)} + \sum_{j=1}^{3} \frac{a_j}{\omega_{0j}^2 - \omega^2 - i\omega\Gamma_j} \quad (2)$$

where the parameters are given in Table 1.

Table 1 The parameters of Equation (2)

| $1+\varepsilon_\infty$ | $\omega_p$(eV) | $\Gamma_0$(eV) | $a_1$ | $\omega_{01}$(eV) | $\Gamma_1$(eV) | $a_2$ | $\omega_{02}$(eV) | $\Gamma_2$(eV) | $a_3$ | $\Gamma_3$(eV) | $\omega_{03}$(eV) |
|---|---|---|---|---|---|---|---|---|---|---|---|
| 2.296 | 9.161 | 0.020 | 12.06 | 5.043 | 0.935 | 27.67 | 6.171 | 1.641 | 5.524 | 4.404 | 0.499 |

**Multipole decomposition.**

The multipole decompositions are accomplished in both the Cartesian basis (source-representation) and the spherical basis (field-representation). The irreducible Cartesian mutipole moments (using the notation of $e^{-i\omega t}$ for electromagnetic waves) are evaluated by[49-52, 68]

$$P_\alpha = -\frac{1}{i\omega}\int J_\alpha d^3r$$

$$M_\alpha = \frac{1}{2c}\int [\mathbf{r}\times\mathbf{J}]_\alpha d^3r$$

$$T_\alpha = \frac{1}{10c}\int \left[(\mathbf{r}\cdot\mathbf{j})r_\alpha - 2r^2 J_\alpha\right]d^3r \qquad (3)$$

$$Q^e_{\alpha\beta} = -\frac{1}{i\omega}\int\left[r_\alpha J_\beta + J_\alpha r_\beta - \frac{2}{3}\delta_{\alpha\beta}(\mathbf{r}\cdot\mathbf{J})\right]d^3r$$

$$Q^m_{\alpha\beta} = \frac{1}{3c}\int\left[[\mathbf{r}\times\mathbf{J}]_\alpha r_\beta + r_\alpha[\mathbf{r}\times\mathbf{J}]_\beta\right]d^3r$$

where $\mathbf{P}$ is the electric dipole moment, $\mathbf{M}$ the magnetic dipole moment, $\mathbf{T}$ the toroidal dipole moment, $\mathbf{Q}^e$ the component of electric quadrupole moment, and $\mathbf{Q}^m$ is the magnetic quadrupole moment. The time-averaged scattered powers of the mutipoles can be written as the following summation:

$$I = \frac{1}{8\pi\varepsilon_0}\left(\frac{2\omega^4}{3c^3}|\mathbf{P}|^2 + \frac{2\omega^4}{3c^3}|\mathbf{M}|^2 + \frac{4\omega^5}{3c^4}\text{Im}(\mathbf{P}^*\mathbf{T}) + \frac{2\omega^6}{3c^5}|\mathbf{T}|^2 + \frac{\omega^6}{20c^5}\sum|\mathbf{Q}^e|^2 + \frac{\omega^6}{20c^5}\sum|\mathbf{Q}^m|^2\right) \quad (4)$$

And the scattering efficiency in vacuum is defined as

$$\sigma_{scs} = \frac{C_{scs}}{2R(2L+d)} \qquad (5)$$

where the scattering cross section reads

$$C_{SCS} = \frac{2Z_0 I}{|E_0|^2} \qquad (6)$$

In Equation (6), $Z_0$ is the wave impedance of vacuum and $E_0$ is the amplitude of incident plane wave.

The scattered electric far field from $\mathbf{P}$, $\mathbf{M}$ and $\mathbf{T}$ follows[49, 69]

$$\mathbf{E}_{far} = \frac{k^2}{4\pi\varepsilon_0} \frac{e^{ikr}}{r} \left[ \hat{\mathbf{n}} \times (\mathbf{P} \times \hat{\mathbf{n}}) + (\mathbf{M} \times \hat{\mathbf{n}}) + ik \cdot \hat{\mathbf{n}} \times (\mathbf{T} \times \hat{\mathbf{n}}) \right] \quad (7)$$

where $k$ is the wave vector in the background medium, $\hat{\mathbf{n}}$ is the unit vector denoting the radiation direction. In Figures 3c and 3d, the far field intensity was calculated at the surface of a sphere of $r = 1$ m, origined in the antenna center.

The spherical multipole moments are obtained by calculating the scattering coefficients through the following field projection[57,70]

$$a_{lm} = \frac{(-i)^{l+1} kr}{h_l^{(1)}(kr) E_0 \left[ \pi (2l+1) l(l+1) \right]^{\frac{1}{2}}} \int_0^{2\pi} \int_0^{\pi} Y_{lm}^*(\theta,\phi) \hat{\mathbf{r}} \cdot \mathbf{E}_s(\mathbf{r}) \sin(\theta) d\theta d\phi$$

$$b_{lm} = \frac{(-i)^l \eta kr}{h_l^{(1)}(kr) E_0 \left[ \pi (2l+1) l(l+1) \right]^{\frac{1}{2}}} \int_0^{2\pi} \int_0^{\pi} Y_{lm}^*(\theta,\phi) \hat{\mathbf{r}} \cdot \mathbf{H}_s(\mathbf{r}) \sin(\theta) d\theta d\phi \quad (8)$$

where $Y_{lm}$ and $h_l^{(1)}$ are the scalar spherical harmonics and the spherical Hankel function of the first kind, respectively; $\mathbf{E}_s$ ($\mathbf{H}_s$) is the scattered electric field (magnetic field), $E_0$ is the amplitude of the incident wave, and $\hat{\mathbf{r}}$ is the unit directional vector. Then the scattering cross sections by the spherical dipoles $a_1$ and $b_1$ read:

$$C_{SCS} = \frac{3\pi}{k^2} \sum_{m=-1}^{1} \left[ |a_{1m}|^2 + |b_{1m}|^2 \right] \quad (9)$$

**Theoretical resonance frequency of the cavity modes.**

In cylindrical coordinate, the field of the gap SPPs at the dielectric layer has the ansatz form:[42,45,63]

$$E_z(\rho,\phi,z) = a(z) J_m(k_{gsp}\rho) e^{im\phi} \quad (10)$$

where $k_{gsp}$ is the wave vector of the gap SPPs in the MIM structure, $a(z)$ the mode profile in $z$ direction, and $m$ is the azimuthal number. At the dielectric layer, the electric field of the gap SPPs at the lateral edge reaches the locale maximum, i.e., the Neumann boundary condition $\partial E_z(R_{eff})/\partial \rho = 0$ shall be satisfied. Combing with equation (10), the resonance condition then reads

$$k_{gsp}(\omega) = \frac{\chi_{mn}}{R_{eff}} \tag{11}$$

where $\chi_{mn}$ denoting the $n$-th root of the first-order derivation of the $m$-th Bessel function, $R_{eff}$ is the appropriate effective radius depending on how strong the field leaks out the circumferences. After obtaining the dispersion relation of the gap SPPs $k_{gsp}(\omega)$, the resonance frequency can be extracted from equation (11). In this study, $k_{gsp}(\omega)$ is obtained from the divergence of the reflection coefficient calculated by the transfer matrix method of a 2D metal-dielectric-metal three layer system.[42,63] The black dashed lines in Figure 5 obtained in this way match the numerical CM bands well by applying $R_{eff} \approx R$, meaning extremely weak leakage of the cavity mode fields.

**ACKNOWLEDGEMENT**

Stimulating discussion with C. T. Chan and Y. T. Chen is greatly appreciated. This work was supported by the NSFC (Grant Nos. 11274083, 11304038, 11374223), the Shenzhen Oversea Talent Plan (No. KQCX20120801093710373), the National Basic Research Program under Grant No. 2012CB921501, and the Project Funded by the Priority Academic Program Development of Jiangsu Higher Education Institutions. We acknowledge helps from the National Supercomputer Shenzhen Center.

**SUPPORTING INFORMATION AVALIABLE**

Absorption spectrum and theoretical prediction of the cavity mode resonances; Scattering spectrum for the other two configurations and radiation decay rate for a nearby dipole; Backward scattering suppression by comparable electric and magnetic dipole moments; Evolution of the scattering spectrum for different gap distance with various loaded material. The Supporting Information is available free of charge via the Internet at http://pubs.acs.org.

Supporting Information of

"Coexistence of Scattering Enhancement and Suppression by Plasmonic Cavity Modes in Loaded Dimer Gap-Antennas"

Qiang Zhang,[†] Jun-Jun Xiao,[†,*] Meili Li,[†] Dezhuan Han,[‡] and Lei Gao[§]

[†]College of Electronic and Information Engineering, Shenzhen Graduate School, Harbin Institute of Technology, Shenzhen 518055, China

[‡]Department of Applied Physics, Chongqing University, Chongqing 400044, China

[§]Jiangsu Key Laboratory of Thin Films, School of Physical Science and Technology, Soochow University, Suzhou 215006, China

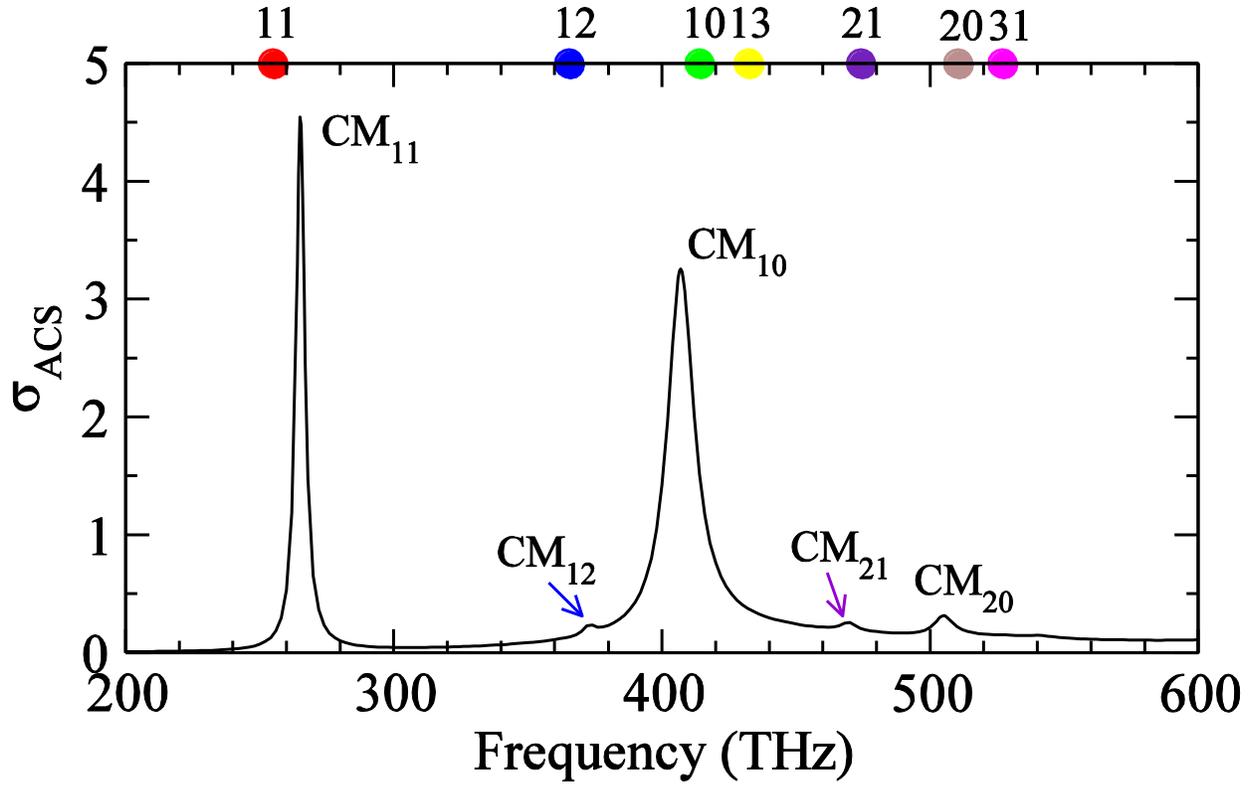

**Figure S1.** The absorption spectrum $\sigma_{ACS}$ of the PDA studied in Figure 2. The color dots on the top $x$-axis mark the theoretically predicted resonance position of the seven CMs labeled with radial and azimuthal numbers '$nm$'. The blue arrow and the indigo arrow indicate the peaks of $CM_{12}$ and $CM_{21}$ resonances, respectively. The remaining $CM_{13}$ and $CM_{31}$ resonance are too weak to have visible features even in the absorption spectrum. Note that all these CMs can be excited efficiently in a circular patch MIM resonator, partially due to the absence of the AM in the same frequency range [1].

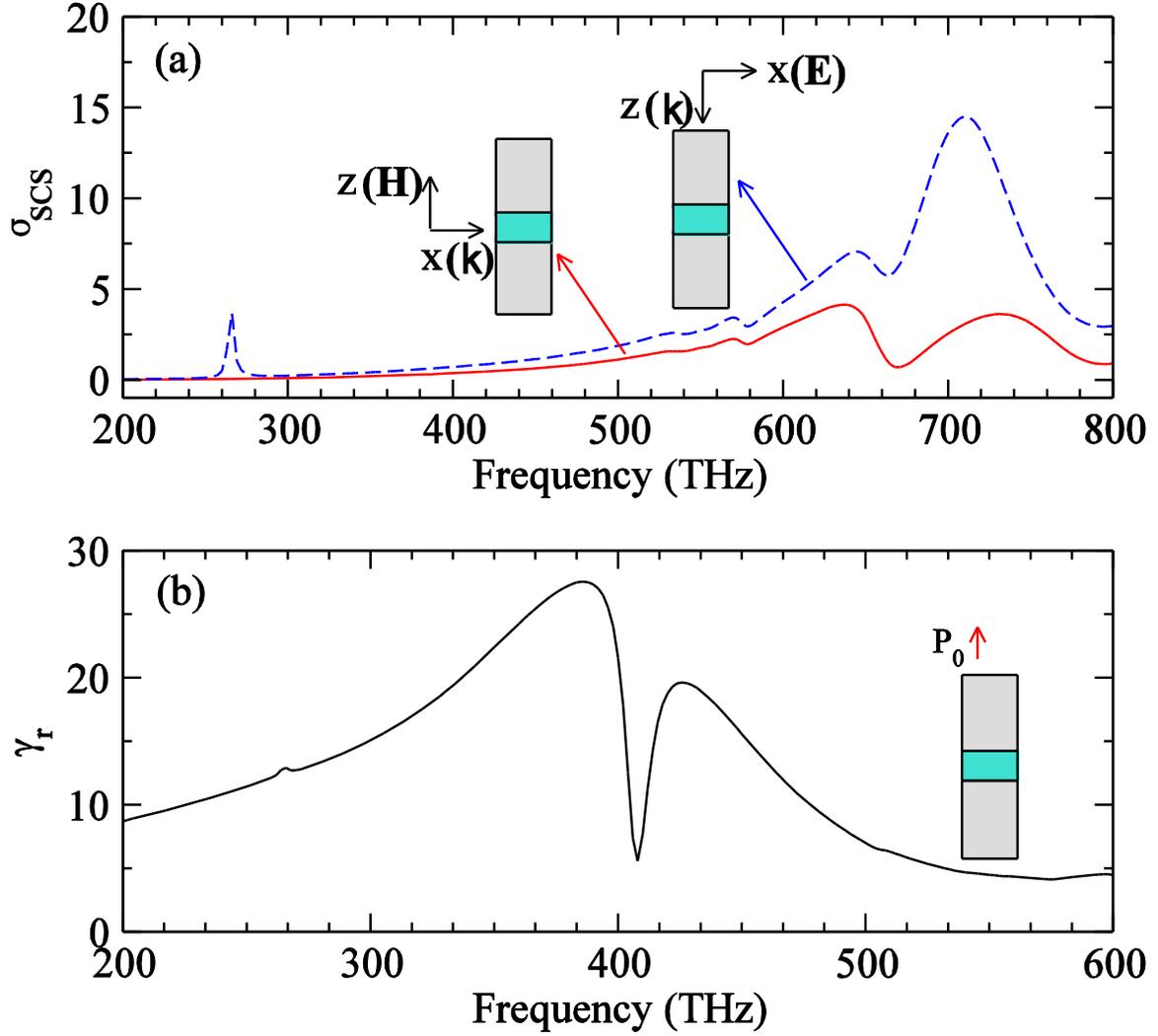

**Figure S2.** (a) The scattering spectrum $\sigma_{SCS}$ of the PDA excited by an incoming plane wave in the other two configurations (see insets): (i) $k$ along the $x$-axis, **H** along the $z$-axis (red solid line) and (ii) $k$ along the $z$-axis, **E** along the $x$-axis (blue dashed line). The longitudinal electric AM is not excitable in these two configurations and the most prominent scattering peak at the high frequency is actually associated with the transverse electric dipole mode [2]. (b) The radiation decay rate $\gamma_r$ [3,4] of an active electric dipole placed near the end of the PDA (the red arrow of the insert). Obviously, the radiation of the dipole is strongly modulated near the cumulative peak and the suppressed dip.

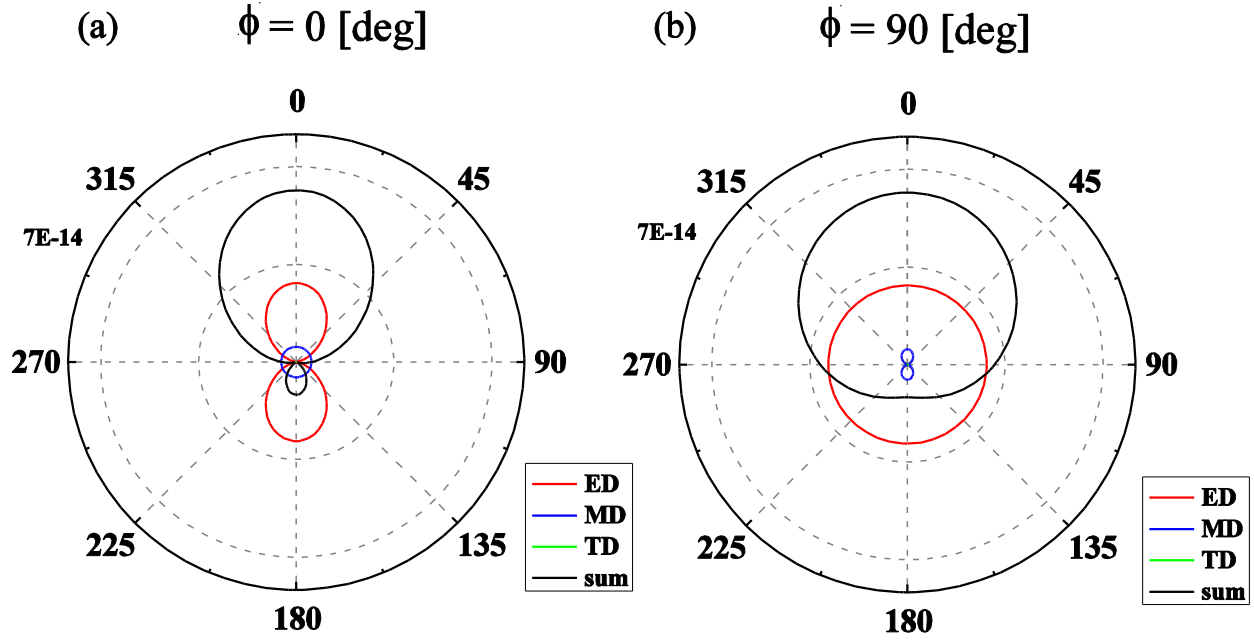

**Figure S3.** The radiation pattern $|\mathbf{E}_{far}(\theta,\phi)|^2$ at the plane of (a) $\phi = 0°$ and (b) $\phi = 90°$ for a PDA with $L = 90$ nm, $R = 50$ nm, $d = 20$ nm, and $\varepsilon_{load} = 3$. The resonant frequency is $f = 462$ THz (i.e., the scattering peak shown in Figure 1c). At this frequency, the $CM_{11}$ and AM both resonantly contribute to the scattered far field (red and blue lines), but with different magnitudes. Their vectorial summation (black lines) shows that the backward scattering is substantially suppressed, although not completely vanishing.

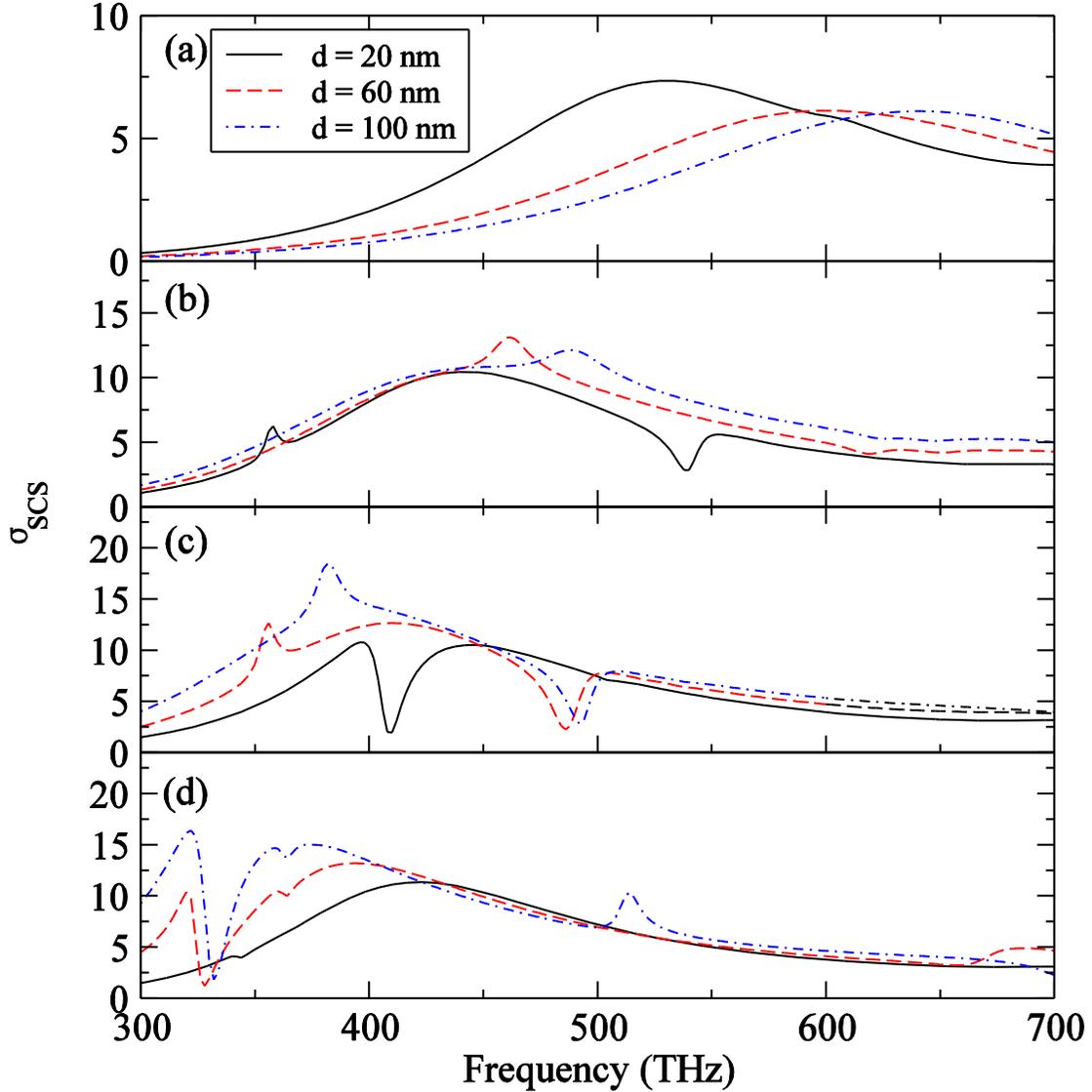

**Figure S4.** Evolutions of the $\sigma_{scs}$ spectrum for different gap distance $d$ of the PDAs with (a) $\varepsilon_{load}=1$, (b) $\varepsilon_{load}=6$, (c) $\varepsilon_{load}=12$, and (d) $\varepsilon_{load}=30$. For $\varepsilon_{load}=1$, the AM is not affected by the CMs and the AM resonance frequency blue shifts for increasing $d$ as predicted by the plasmon ruler equation and the hybridization model [5,6]. When $\varepsilon_{load}$ increases, the CMs are tuned to overlap with the broad AM spectrally and couple with the AM, deteriorating the blue-shift trend. Especially, when the AM is strongly modified by the Fano dip in (c) and (d), its peak is contrarily red-shifted as the separation distance $d$ increases.